\documentclass[fleqn,usenatbib]{mnras}

\usepackage{newtxtext,newtxmath}

\usepackage[T1]{fontenc}

\DeclareRobustCommand{\VAN}[3]{#2}
\let\VANthebibliography\thebibliography
\def\thebibliography{\DeclareRobustCommand{\VAN}[3]{##3}\VANthebibliography}

\usepackage{graphicx}	
\usepackage{amsmath}	

\title[AGN In Low-Mass Dwarfs]{Environmental Dependence of X-Ray Emission From The Least Massive Galaxies}

\author[M. Mićić et al.]{
Marko Mićić,$^{1}$\thanks{E-mail: micic@ou.edu}
Xinyu Dai,$^{1}$
Nick Shumate,$^{1}$
Khoa Nguyen Tran, $^{1}$
and Heechan Yuk$^{1}$
\\
$^{1}$Homer L. Dodge Department of Physics and Astronomy, The University of Oklahoma, Norman, OK 73019, USA\\
}

\date{Accepted XXX. Received YYY; in original form ZZZ}

\pubyear{\the\year{}}

\begin{document}
\label{firstpage}
\pagerange{\pageref{firstpage}--\pageref{lastpage}}
\maketitle

\begin{abstract}
The low-mass end of low-mass galaxies is largely unexplored in AGN studies, but it is essential for extending our understanding of the black hole-galaxy coevolution. We surveyed the 3D-HST catalog and collected a sample of 546 dwarf galaxies with stellar masses log(M$_*$/\(M_\odot\))$<$8.7, residing in the GOODS-South deep field. We then used the unprecedented depth of Chandra available in the GOODS-South field to search for AGN. We carefully investigated the factors that could play roles in the AGN detectability, such as Chandra's point-spread function and the redshift- and off-axis-dependent detection limits. We identified 16 X-ray sources that are likely associated with AGN activity. Next, we evaluated the environment density of each galaxy by computing tidal indices. We uncovered a dramatic impact of the environment on AGN triggering as dwarfs from high-density environments showed an AGN fraction of 22.5\%, while the median stellar mass of this subset of dwarfs is only log(M$_*$/\(M_\odot\))=8.1. In contrast, the low-density environment dwarfs showed an AGN fraction of only 1.4\%, in line with typically reported values from the literature. This highlights the fact that massive central black holes are ubiquitous even at the lowest mass scales and demonstrates the importance of the environment in triggering black hole accretion, as well as the necessity for deep X-ray data and proper evaluation of the X-ray data quality. Alternatively, even if the detected X-ray sources are related to stellar mass accretors rather than AGN, the environmental dependence persists, signaling the impact of the environment on galaxy evolution and star formation processes at the lowest mass scales. Additionally, we stacked the X-ray images of non-detected galaxies from high- and low-density environments, revealing similar trends.
\end{abstract}




\section{Introduction}
The prevailing wisdom suggests that almost all massive galaxies contain a supermassive black hole in their nucleus. Additionally, various scaling relations between central supermassive black holes and galactic properties have been observed, highlighting the intimate coevolution between the two components \citep[e.g.,][]{2013ApJ...764..184M, Reines_2015}. However, it remains unclear whether these relations hold at the lowest mass scales or if central massive black holes are frequent among low-mass dwarf galaxies, significantly limiting our understanding of the earliest stages of galaxy formation and evolution. This issue is particularly noticeable among the low-mass end of low-mass galaxies since such galaxies are usually overlooked due to various technical limitations associated with the detectability of their massive black holes. 

Some potential methods rely on detecting unique signatures of actively accreting massive black holes. For example, the Baldwin-Phillips-Terlevich (BPT) diagram is commonly used to identify the presence of AGN \citep{1981PASP...93....5B}. It utilizes the fact that AGN produces a radiation field that is harder than that from star formation processes, leading to enhanced excitation of nebular lines, and offers a simple method to distinguish between different ionization mechanisms. Unfortunately, this method tends to be inaccurate and unreliable at low-mass scales as it frequently misclassifies dwarfs with clear multiwavelength AGN evidence as star-forming galaxies \citep{10.1093/mnras/staa040}. Furthermore, in shallow surveys, which typically only detect dwarfs with significant star formation activity \citep{2025MNRAS.538..153K}, the star formation can swamp the optical emission lines, further complicating the detection of AGN \citep{2024MNRAS.528.5252M}. The BPT classification of dwarf AGN can be improved through the use of spatially resolved spectroscopy and spaxel-by-spaxel classification methods. The MaNGa AGN dwarf galaxies (MAD) survey used this approach to detect hundreds of dwarf AGN and derive an AGN fraction of $\sim$20\%, significantly higher than that of single-fiber spectroscopy studies \citep{2024MNRAS.528.5252M}. However, the spatially resolved spectroscopy method is limited to relatively nearby dwarfs (the median redshift of AGN-hosting dwarfs from the MAD survey is $z$=0.0036) and focused on massive dwarfs (the median stellar mass of AGN-hosting dwarfs from the MAD survey is log(M$_*$/\(M_\odot\))=9.2). Another approach relies on detecting the radio emission from dwarf galaxies since all black holes, even the weakly accreting ones, are expected to produce a radio continuum at centimeter wavelengths, which is almost immune to dust extinction \citep{2020ApJ...888...36R, Eberhard_2025}. However, this method is significantly limited by the detection limits of existing radio surveys. Even the deepest surveys can only detect the youngest AGN in dwarfs \citep{2022MNRAS.511.4109D}.

Finally, the most promising approach relies on detecting X-ray emission from the accretion disk. Even though dwarf galaxy AGN are expected to be relatively faint in X-rays \citep{2016A&A...596A..64S}, often requiring hundreds of kiloseconds of exposure time to detect them even at modest distances, this method remains the most popular \citep{2015ApJ...805...12L,2016ApJ...831..203P,2018MNRAS.478.2576M,10.1093/mnras/staa040}. All these studies consistently found low AGN fractions, typically gravitating around a few percent. However, their galaxy samples were often skewed towards the high-mass dwarfs, neglecting the low-mass end of the dwarf regime.

In this work, we present the first attempt to fill this gap by utilizing Chandra archival X-ray data to investigate the presence of AGN among the least massive galaxies known. We investigated the AGN fractions and the mechanisms that trigger AGN. The paper is organized as follows: in Section \ref{sec2} we present the data used in the paper, methodology, and galaxy and AGN selection criteria; in Section \ref{sec3} we present our results, AGN fractions, and triggering mechanisms; in Section \ref{sec4} we present environment-dependent stacking analysis. The uncertainties quoted throughout the paper are obtained from the Poissonian statistics at 1$\sigma$ levels \citep{1986ApJ...303..336G}, unless otherwise specified.



\section{DATA AND METHODOLOGY}\label{sec2}
\subsection{Dwarf Galaxy Sample}
We used data derived from the 3D-HST survey, a 248-orbit HST Treasury program \citep[M16 and S14, hereafter]{2016ApJS..225...27M, 2014ApJS..214...24S}. The survey performed WFC3/G141 grism spectroscopy across five CANDELS deep fields, providing information on distances, stellar masses, star formation rates, emission line measurements, and stellar ages for over 100,000 galaxies. Due to the extreme sensitivity of available X-ray data, we concentrated our work on the subset of galaxies residing in the GOODS-South field. More details about this can be found in Section \ref{susbet2.2}.

Since all derived galactic properties depend on distance, we pay special attention to the redshift quality analysis. The 3D-HST survey catalog provides three types of redshifts: (1) ground-based spectroscopic redshift, where available; (2) photometric redshifts from S14 obtained by fitting the spectral energy distributions with a linear combination of seven galaxy templates using the EAZY code \citep{Brammer_2008}; (3) grism redshifts from M16, obtained by simultaneously fitting the 2D spectra and multi-band photometry.
\begin{figure}
	\includegraphics[width=\columnwidth]{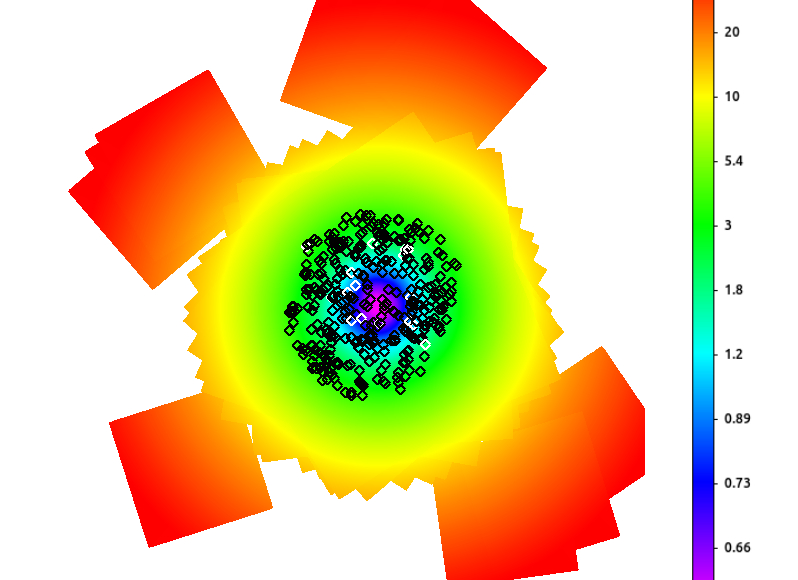}
    \caption{Chandra point-spread function map of the GOODS-South field. The black diamonds show the locations of dwarf galaxies from our sample, highlighting the absence of galaxies located far off-axis. The white diamonds show locations of dwarf galaxies with detected X-ray emission.}
    \label{fig:fig1}
\end{figure}
\begin{figure}
	\includegraphics[width=\columnwidth]{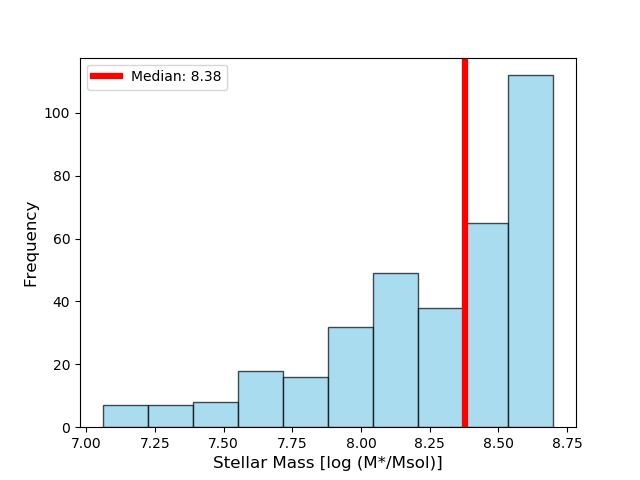}
    \caption{The stellar mass distribution of dwarf galaxies. The red vertical line shows the median stellar mass value.}
    \label{fig:fig2}
\end{figure}
\begin{figure}
	\includegraphics[width=\columnwidth]{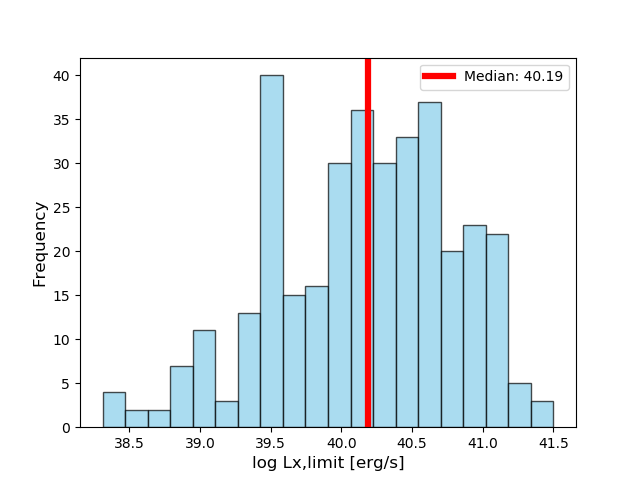}
    \caption{Distribution of detection limits in our sample. The red vertical line shows the median detection limit value.}
    \label{fig:fig3}
\end{figure}
We keep all galaxies with available ground-based spectroscopic redshifts, as their distances are calculated with great precision, although the number of such galaxies is relatively small. Regarding the rest of the galaxies from the GOODS-South field, we compared the S14 and M16 redshifts, analyzed uncertainties in the redshift measurements, and kept only those galaxies with consistent and well-constrained redshifts. The first requirement was that the redshifts obtained by two different independent methods do not diverge significantly: |z$_{S14}$-z$_{M16}$|$<$0.05, where z$_{S14}$ and z$_{M16}$ are the redshifts obtained by S14 and M16, respectively. This filter eliminated galaxies with inconsistent, potentially erroneous redshift measurements. The typical uncertainty of 3D-HST redshifts is 0.003 $\times$ (1+$z$), but it varies on an individual basis since it depends on redshift, brightness, and star formation rate. We manually examined the redshifts of all candidate dwarfs and included only those with well-constrained redshifts, within expected uncertainties. The redshift variations across a relatively narrow range of values would not significantly impact the derivations of galactic properties, including the key property of stellar mass.

The final cut only included galaxies with stellar masses of M$_*<$ 5$\times$10$^8$ \(M_\odot\). This stellar mass limit is an order of magnitude lower than what is typically used in similar dwarf galaxy works from the literature, highlighting our focus on the low-mass end of low-mass galaxies. The stellar masses were adopted from the 3D-HST catalog. They were derived using the FAST code, assuming the \cite{2003MNRAS.344.1000B} stellar population synthesis model, \cite{2003PASP..115..763C} initial mass function, solar metallicity, and exponentially declining star formation histories. We refer the reader to \cite{2014ApJS..214...24S} for more detailed information. After applying all the filters, we constructed a parent sample of 546 galaxies with well-constrained redshifts and stellar masses likely in the low-mass end of dwarf galaxies.
\subsection{Chandra X-ray data}\label{susbet2.2}
GOODS-South field has been observed with the Chandra X-ray telescope 111 times for a total exposure time of $\sim$7 Msec. It is the longest exposure of any field ever observed with Chandra and offers an opportunity to study X-ray sources at unprecedented depths. X-ray sources associated with low-mass dwarf galaxies are expected to be faint, and their detectability depends on the off-axis angle. The farther the source is from the aim point of the observation, the more diffuse it will appear, particularly affecting the detectability of faint, low-count-rate sources. For this reason, we constructed the point spread function (PSF) map of the GOODS-South field and filtered our parent sample to keep only galaxies with a PSF less than 4\arcsec. Our refined sample of low-mass dwarfs with confident redshift measurements and deep and high-quality X-ray data contains 352 galaxies. The PSF map of the GOODS-South field with overlaid locations of low-mass dwarfs is shown in Figure \ref{fig:fig1}. The stellar mass distribution of the sample is shown in Figure \ref{fig:fig2} and the median stellar mass is log(M$_*$/\(M_\odot\))=8.38. The dwarfs span a range of redshifts from 0.05 to 1, with a median value of 0.34.

The flux sensitivity of the Chandra GOODS-South survey peaks at the center of the observation and declines with increasing distance from the center \citep{2011ApJS..195...10X}. We obtained the flux sensitivity limits for each dwarf galaxy from our sample by calculating their off-axis angles. The impact of redshifts must be accounted for to convert the flux limits into luminosity limits and find the faintest X-ray source possible to see in each galaxy. Figure \ref{fig:fig3} shows the luminosity limit distribution histogram, and the median value is log L$_x$=40.19. Although the redshifts in our sample span a wide range of values, we find that low-luminosity X-ray sources are detectable to some extent across the entire sample.
\subsection{AGN Selection Criteria}
\label{subsec2.3}
We then crossmatched 352 dwarf galaxies from our sample with available catalogs of Chandra-detected X-ray sources from the GOODS-South field, looking for X-ray sources within 1\arcsec \citep{2011ApJS..195...10X, 2014MNRAS.443.1999B,2016ApJ...823...95C,2017ApJS..228....2L}. We found 16 matches with X-ray luminosities ranging from log L$_X$=39.7 erg s$^{-1}$ to log L$_X$=42.4 erg s$^{-1}$, with the median value of L$_X$=40.2 erg s$^{-1}$. We then evaluated the contribution of unresolved populations of X-ray binaries (XRBs), powered by stellar-mass accretors, to the observed X-ray emission, by using the scaling relation from \cite{Lehmer_2010}:
\begin{equation}
    L_{\textup{XRB}}=\alpha_0M_*+\beta_0SFR
\end{equation}
in erg s$^{-1}$, where $\alpha_0$=(9.05$\pm$0.37)$\times$10$^{28}$ erg s$^{-1}$ \(M_\odot\)$^{-1}$, $\beta_{0}$=(1.62$\pm$0.22)$\times$10$^{39}$ erg s$^{-1}$ (\(M_\odot\) yr$^{-1}$)$^{-1}$, M$_*$ is stellar mass, and SFR is star formation rate in units of \(\textup{M}_\odot\) yr$^{-1}$. The  M$_*$ term is proportional to low-mass XRBs, while the SFR term is proportional to high-mass XRBs. We found that XRB contributions account for 0.02 to 17.3\% of the observed X-ray emission, with a median value of 3\%. The observed X-ray emission in all 16 X-ray sources exceeds the expected XRB contributions by a factor of at least three, a cutoff typically used in the literature as an AGN diagnostic \citep{2024MNRAS.527.1962B}.

The hot interstellar medium gas can also produce substantial X-ray emission in some cases. We used the relation from \cite{2012MNRAS.426.1870M} to estimate this contribution:
\begin{equation}
    L_{\textup{hot}}=(8.3\pm0.01)\times10^{38}SFR
\end{equation}
and found that in all 16 cases, it accounts for $<$6\% of the observed X-ray emission, with a median value of only 1\%.

Another potential explanation for the observed X-ray emission is ultraluminous X-ray sources (ULXs). ULXs are powered by stellar-mass black holes or neutron stars undergoing supercritical accretion \citep{beg2002ApJ...568L..97B} or having geometrically beamed emission \citep{king2001ApJ...552L.109K} and exceeding theoretical maximum luminosity. However, ULXs with X-ray luminosity exceeding 10$^{40}$ erg s$^{-1}$ (the median X-ray luminosity of our sample is 1.6$\times$10$^{40}$ erg s$^{-1}$) are quite rare. The Milky Way has none, and all local group dwarfs combined have none, likely due to insufficient stellar mass \citep{Swartz_2008}. \cite{kov2020MNRAS.498.4790K} studied the local Universe ULXs and found that 0.5 ULXs with luminosity exceeding 10$^{40}$ erg s$^{-1}$ are expected to be found per 10$^{12}$ \(M_\odot\) of stellar mass. This implies that one ULX is expected to be found in 10,000 dwarf galaxies with a stellar mass equal to the median stellar mass of our sample Even though we cannot conclusively rule out the possibility that at least some of the observed X-ray sources are XRBs or ULXs, we conclude, based on the available data, that all 16 X-ray sources require an additional intensive source of X-ray emission, i.e., likely an AGN.

\section{AGN Fractions and triggering mechanisms}\label{sec3}
\begin{figure}
	\includegraphics[width=\columnwidth]{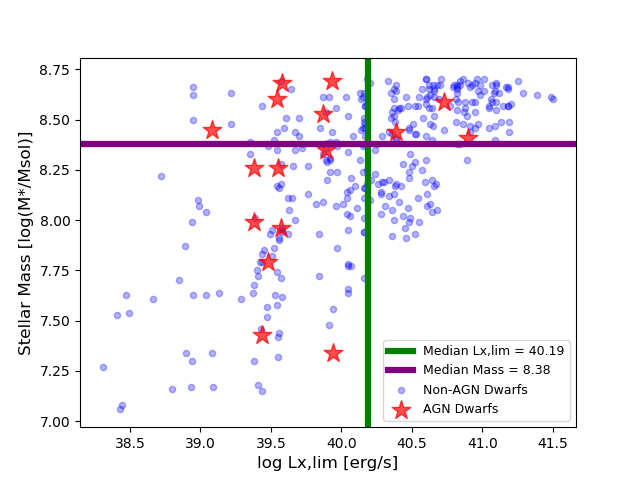}
    \caption{Distribution of AGN (red stars) and non-AGN (blue dots) dwarfs in the stellar mass-detection limit space. The green vertical line represents the median detection limit, while the purple horizontal line represents the median stellar mass of the sample.}
    \label{fig:fig4}
\end{figure}
\begin{figure}
	\includegraphics[width=\columnwidth]{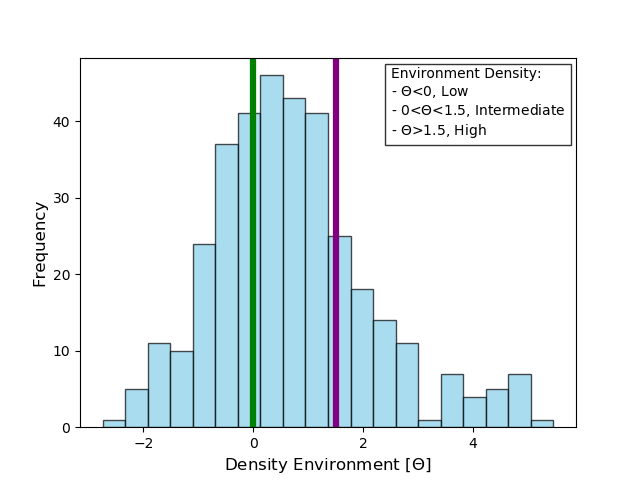}
    \caption{Distribution of tidal indices across our dwarf sample. The region to the left of the green vertical line denotes the low-density environment, the region between the two vertical lines denotes the intermediate-density environment, and the region to the right of the purple vertical line denotes the high-density environment \citep{pe10.1093/mnras/stw757}.}
    \label{fig:fig5}
\end{figure}
\begin{figure}
	\includegraphics[width=\columnwidth]{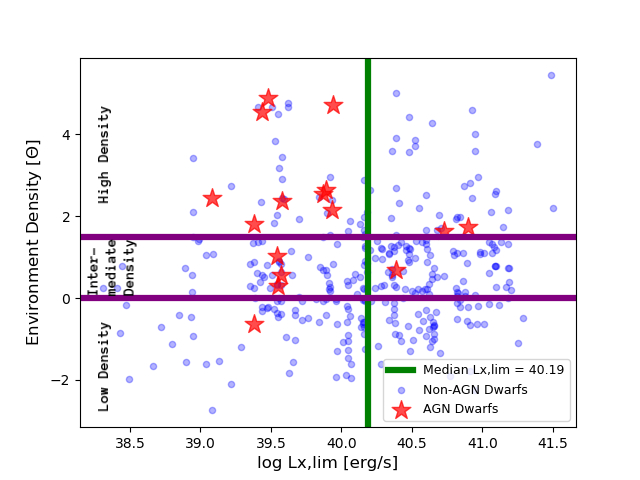}
    \caption{Distribution of AGN (red stars) and non-AGN (blue dots) dwarfs in the environment density-detection limit space.}
    \label{fig:fig6}
\end{figure}
\begin{figure}
	\includegraphics[width=\columnwidth]{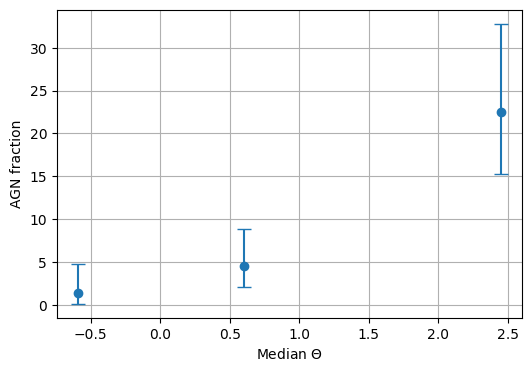}
    \caption{AGN fraction in the three types of environment. Higher $\Theta$ means a higher density environment.}
    \label{fig:fig7}
\end{figure}
Figure \ref{fig:fig4} shows the distribution of AGN and non-AGN dwarfs in the stellar mass-detection limit space. We notice a stark difference between the higher and lower detection limits, that is, between the shallower and deeper portions of the sample. The shallower half contains only three AGN, and the implied AGN fraction is 1.7\%$^{+1.6}_{-0.9}$. However, the deeper half contains thirteen AGN, and the implied AGN fraction is 7.4\%$^{+2.6}_{-2.1}$. Although this is not unexpected, it underscores the importance of obtaining deep X-ray data when working with low-mass dwarf galaxies. This significant difference confirms findings by \cite{2016A&A...596A..64S} that dwarf galaxies with clear AGN optical signatures often lack associated strong X-ray emission, which was explained by weak accretion, obscuration of accretion disk along the line of sight, or unorthodox photon production processes. On the other hand, we do not notice any stellar mass evolution. The high and low mass halves of the sample contain eight AGN each, and the implied fraction of AGN is 4.5\%$^{+2.3}_{-1.5}$.

Various theoretical, computational, and observational studies have suggested that galaxy interactions may be a significant trigger of black hole accretion. However, the impact of the environment on the presence of AGN in dwarf galaxies has been poorly explored. Cosmological simulations suggested that minor mergers, when a dwarf galaxy interacts with a much larger companion, could initiate bursts of violent, short-lived accretion and lead to an increase of black hole mass by up to ten times, posing as potentially an essential channel for black hole growth \citep{Callegari_2011,cap10.1093/mnras/stu2500,2012ApJ...748L...7V}. This claim is not supported by observations since only a few secondary AGN in minor mergers have been discovered \citep{Koss_2012, Comerford_2015, Secrest_2017, Liu_2018, Mi_i__2024}. Additionally, a few AGN and dual AGN have been discovered in ongoing dwarf-dwarf galaxy mergers \citep{2014ApJ...787L..30R,2021ApJ...912...89K, 
 Kimbrell_2024,2023ApJ...944..160M}. Finally, \cite{2024ApJ...968L..21M} conducted the largest survey of AGN in dwarf-dwarf galaxy mergers, discovering six new sources and concluding that dwarf-dwarf galaxy pairs are six to ten times more likely to have an AGN than isolated dwarf galaxies of the same stellar mass and at the same redshift.

We quantified the environment of all galaxies in our sample by calculating the tidal indices, $\Theta$, to test the impact of the environment on the triggering of AGN among the low-mass end of dwarf galaxies. The tidal index is a measure of the density of the environment introduced by \cite{1998A&A...331..891K}, used in various works \citep{2011ApJ...743....8W, pe10.1093/mnras/stw757}, and defined as:
\begin{equation}
    \Theta=log_{10}(\frac{M_*[\times10^{11}M_{\odot}]}{D^3_{project}[Mpc]})
\end{equation}
where M$_*$ and D$_{project}$ are the stellar mass and projected distance of the neighbor galaxy. We searched for neighbors of any mass of our dwarf galaxies within $\Delta$z=0.02, for which the tidal index $\Theta$ reaches the maximum value. A higher tidal index denotes a denser environment. Galaxies with a negative tidal index are considered isolated. In contrast, those with $\Theta$ greater than 1.5 are considered to reside in dense environments, that is, likely in the process of active interaction with a massive neighbor. In Figure \ref{fig:fig5}, we show the distribution of the tidal indices in our sample. The leftmost portion of the histogram denotes low-density environments, the middle portion denotes intermediate density, and the far right denotes high-density environments. The boundary values of $\Theta$ between different types of environments are adopted from \cite{pe10.1093/mnras/stw757}. Most dwarfs reside in low-density environments, 120/352, or 34\%, and intermediate-density environments, 153/352, or 43\%, while only 79/352 dwarfs, or 32\% reside in high-density environments. 

The high-density environment dwarfs contain eleven potential AGN, or 69\% of the total number of AGN. Intermediate-density dwarfs contain four AGN, or 25\%, while low-density dwarfs contain only one AGN, or 6\%. The distribution of AGN and non-AGN galaxies with respect to their tidal index is shown in Figure \ref{fig:fig6}. Furthermore, if only the portion of the sample with higher-quality X-ray data and deeper detection limits is considered, we find that the AGN fraction in dwarfs residing in high-density environments is 22.5\%$^{+10.3}_{-7.2}$. In comparison, the AGN fractions in dwarfs residing in intermediate- and low-density environments are 4.5\%$^{+4.3}_{-2.4}$ and 1.4\%$^{+3.4}_{-1.2}$, respectively. The results are graphically represented in Figure \ref{fig:fig7}. The AGN fraction of the high-density dwarfs is far higher than in any similar work that used X-ray data to search for dwarf AGN. Additionally, the median stellar mass of the high-density and deep detection limits subsample is only log(M$_*$/\(M_\odot\))=8.1, demonstrating the dramatic impact of the environment on AGN triggering in the low-mass end of dwarf galaxies, the necessity of deep high-quality X-ray observations, and proper characterization of off-axis angle, point-spread function, and redshift on detection limits.

However, due to 3D-HST redshift uncertainties, we can only assess projected distances, rather than actual 3D distances. This does not affect the low-density environment subset, as there are no potential galaxy neighbors that could cause an increased tidal index value in either the 3D or projected distance scenario. On the other hand, the high-density environment subset is heavily impacted because the $\Delta$z=0.02 variation in redshift can lead to a dramatic change in the tidal index when 3D distance is considered. Therefore, the high-density environment subset is likely comprised of a mixture of low-density interlopers and genuine high-density galaxies. The low-density interlopers are identical to low-density environment dwarfs in terms of stellar mass, redshift, and detection limits, with the only difference being erroneous estimates of tidal indices of the former due to insufficient resolution. As such, they are likely to have comparable AGN fractions, which we previously determined to be $\sim$1.4\%. However, a mixture of low-density interlopers and genuine high-density dwarfs shows a significantly elevated AGN fraction of $\sim$22.5\%, suggesting that this value is a lower limit of the AGN fraction for the genuine high-density environment dwarfs subset.

As discussed in Section \ref{subsec2.3}, there is a non-negligible probability that some of the X-ray sources are unresolved populations of XRBs or ULXs instead of AGN. However, the environmental dependence of observed X-ray emission persists. The increased presence of XRBs and ULXs in a galaxy signals elevated activity of star formation processes. This implies that, in a non-AGN scenario, galaxy interactions and mergers trigger star formation in low-mass dwarf galaxies. Indeed, computational works have found that galaxy interactions that do not result in a merger (i.e., flybys) cause enhancements in star formation and contribute to approximately 10\% of stellar mass formation in the dwarf regime \citep{2021MNRAS.500.4937M}. Similarly, observations suggest that $\sim$20\% of star formation in dwarfs is triggered by galaxy interactions, independent of the morphological type \citep{2024MNRAS.529..499L, 2024MNRAS.533.3771L}. Additionally, \cite{2020AJ....159..103K} studied isolated dwarf galaxies with tidal debris, the derivative of dwarf-dwarf mergers, and found that they show higher star formation rates and have bluer colors than their unperturbed counterparts. More recent works also suggest that dwarf-dwarf interactions profoundly affect star formation as they can both trigger and quench star formation. However, quenching is believed to be temporary \citep{2024ApJ...963...37K}. In conclusion, in the non-AGN scenario, our findings that dwarfs residing in high-density environments have elevated star formation are in accordance with similar works from the literature.
\section{Stacking Analysis}\label{sec4}
\begin{figure}
	\includegraphics[width=\columnwidth]{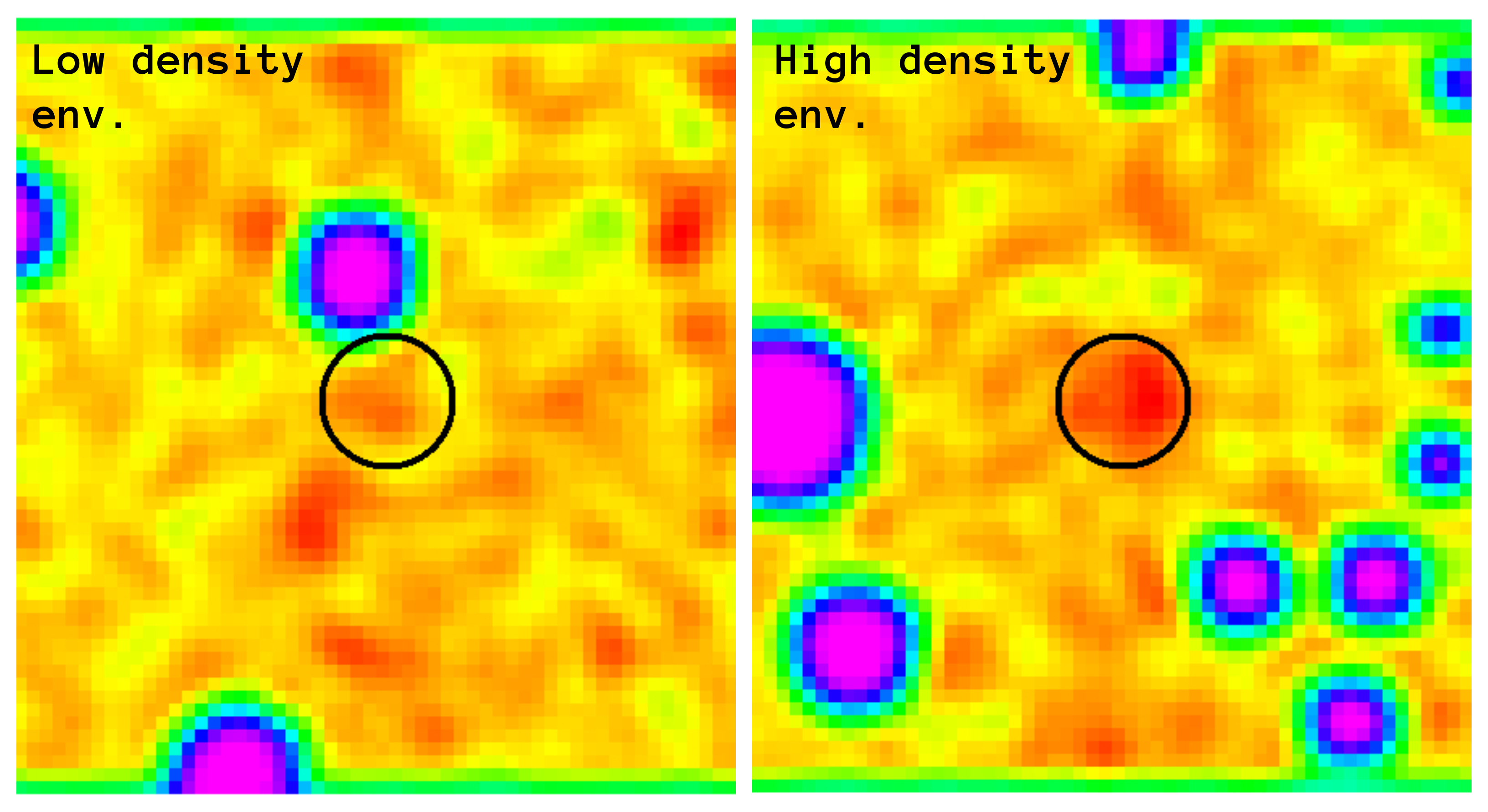}
    \caption{Left: Stacked image of 31 non-detected dwarfs from low-density environments with deep detection limits. Right: Stacked image of 31 non-detected dwarfs from low-density environments with deep detection limits. The purple regions are locations of real sources that have been masked. The redder color corresponds to increased X-ray emission. The black circles are central 2\farcs5 circular apertures placed at the location of stacked galaxies.}
    \label{fig:fig8}
\end{figure}
We perform an additional test by stacking the X-ray images of non-detected galaxies in high- and low-density environments. We began by downloading, reprocessing, and merging all appropriate Chandra observations using the Chandra Interactive Analysis of Observations (CIAO 4.17). We then cut out small, 30$\times$30$\arcsec$ images centered on each of the 31 high-density environment dwarfs with deep detection limits but without an X-ray source. Stacking these 31 cut-out images allowed us to further reduce the detection limits and create an X-ray image with $\approx$200 Msec of exposure time. The same procedure was repeated for 31 low-density environment dwarfs with deep detection limits and no X-ray sources. The two subsets of dwarfs are identical in terms of stellar mass and detection limits, with the only difference being the environment. However, the high-density environment stack reveals a statistically significant detection of 323$\pm$79 net photons within the central 2$\farcs$5 circular aperture. On the other hand, the low-density environment stack shows an enhancement of 91$\pm$78 net photons. This result reinforces the previous findings as it demonstrates the environment-related production of X-ray emission among the least massive galaxies. Both stacked images are shown in Figure \ref{fig:fig8}.
\section{Summary}
In this paper, we: 
\begin{itemize}
    \item Constructed a large sample of very low-mass dwarf galaxies (M$_*<$ 5$\times$10$^8$ \(M_\odot\))
    \item Constructed the PSF maps and calculated off-axis angle- and redshift-dependent detection limits to further refine the dwarf galaxy sample and evaluate the quality of the X-ray data
    \item Found 16 X-ray detections likely due to the AGN activity, but at least some of the sources may be produced by stellar mass accretors, unresolved populations of XRBs, or ULXs
    \item Calculated tidal indices, $\Theta$, for all galaxies, quantifying the density of their environments
    \item Uncovered a major impact of the environment on enhanced X-ray emission and, likely, AGN activation: 22.5\% high-density environment dwarfs showed detectable X-ray emission, compared to only 1.4\% low-density environment dwarfs
    \item Found the same pattern among stacked subsamples of non-detected high- and low-density environment dwarfs. 
\end{itemize}

\section*{Acknowledgements}

MM is supported by the University of Oklahoma's Department of Physics and Astronomy Dodge Family Postdoctoral Fellowship. The authors thank Sugata Kaviraj for helpful discussions.




\bibliographystyle{mnras}
\bibliography{example} 





\bsp	
\label{lastpage}
\end{document}